\documentclass[aps,prb,superscriptaddress,showkeys,twocolumn]{revtex4}

\usepackage{amsmath,amssymb}
\usepackage{amsfonts} 	
\usepackage{graphicx}   
\usepackage{verbatim}   
\usepackage{color}      
\usepackage{subfigure}  
\usepackage[normalem]{ulem} 
\usepackage[super]{nth} 
\usepackage{pbox}

\begin{document}

\title{The local density of optical states in the 3D band gap of a  finite photonic crystal}

\author{Charalampos P. Mavidis}
\email{mavidis@iesl.forth.gr}
\affiliation{Department of Materials Science and Technology, University of Crete, Heraklion, Crete, Greece}
\affiliation{Institute of Electronic Structure and Laser, Foundation for Research and Technology Hellas, N. Plastira 100, 70013 Heraklion, Crete, Greece}

\author{Anna C. Tasolamprou}
\affiliation{Institute of Electronic Structure and Laser, Foundation for Research and Technology Hellas, N. Plastira 100, 70013 Heraklion, Crete, Greece}

\author{Shakeeb B. Hasan}
\affiliation{Complex Photonic Systems (COPS), MESA+ Institute for Nanotechnology, University of Twente, P.O. Box 217, 7500 AE Enschede, The Netherlands}
\altaffiliation{Current address: ASML Netherlands B.V., 5504 DR Veldhoven, The Netherlands}

\author{Thomas Koschny}
\affiliation{Ames Laboratory and Department of Physics and Astronomy, Iowa State University, Ames, Iowa 50011, USA}

\author{Eleftherios N. Economou}
\affiliation{Institute of Electronic Structure and Laser, Foundation for Research and Technology Hellas, N. Plastira 100, 70013 Heraklion, Crete, Greece}
\affiliation{Department of Physics, University of Crete, Heraklion, Greece}

\author{Maria Kafesaki}
\affiliation{Department of Materials Science and Technology, University of Crete, Heraklion, Crete, Greece}
\affiliation{Institute of Electronic Structure and Laser, Foundation for Research and Technology Hellas, N. Plastira 100, 70013 Heraklion, Crete, Greece}

\author{Costas M. Soukoulis}
\affiliation{Institute of Electronic Structure and Laser, Foundation for Research and Technology Hellas, N. Plastira 100, 70013 Heraklion, Crete, Greece}
\affiliation{Ames Laboratory and Department of Physics and Astronomy, Iowa State University, Ames, Iowa 50011, USA}

\author{Willem L. Vos}
\email{w.l.vos@utwente.nl}
\affiliation{Complex Photonic Systems (COPS), MESA+ Institute for Nanotechnology, University of Twente, P.O. Box 217, 7500 AE Enschede, The Netherlands}

\begin{abstract}
A three-dimensional (3D) photonic band gap crystal is an ideal tool to completely inhibit the local density of optical states (LDOS) at every position in the crystal throughout the band gap. 
This notion, however, pertains to ideal infinite crystals, whereas any real crystal device is necessarily finite. 
This raises the question as to how the LDOS in the gap depends on the position and orientation inside a finite-size crystal. 
Therefore, we employ rigorous numerical calculations using finite-difference time-domain (FDTD) simulations of 3D silicon inverse woodpile crystals filled with air or with toluene, as previously studied in experiments.
We find that the LDOS versus position decreases exponentially into the bulk of the crystal. 
From the dependence on dipole orientation, we infer that the characteristic LDOS decay length $\ell_{\rho}$ is mostly related to far-field dipolar radiation effects, whereas the prefactor is mostly related to near-field dipolar effects. 
The LDOS decay length has a remarkably similar magnitude as the Bragg length for directional transport, which suggests that the LDOS in the crystal is dominated by vacuum states that tunnel from the closest interface towards the position of interest. 
Our work leads to design rules for applications of 3D photonic band gaps in emission control and lighting, quantum information processing, and in photovoltaics. 
\end{abstract}


\maketitle	

\section{Introduction}
\label{sec:intro} 
Controlling the properties of matter by means of quantum light lies at the heart of quantum optics and cavity quantum electrodynamics (cQED). 
A prime example is the control of the radiative rate of elementary emitters such as atoms, ions, molecules, or quantum dots. 
Such control is essential for myriad applications ranging from miniature lasers and light-emitting diodes~\cite{Yablonovitch1987PRL,Tandaechanurat2011NP}, via single-photon sources for quantum information processing~\cite{Barnes2002EPJD}, to solar energy harvesting~\cite{Fleming2002N}. 
To explore such new applications, a suitably tailored dielectric environment is required wherein the  vacuum fluctuations, that play a central role in spontaneous emission~\cite{Haroche1992Book,Milonni1994Book}, are controlled. 
Much after the early realization by Purcell~\cite{Purcell1946PR} that an emitter's environment such as a cavity controls the emission rate, spontaneous emission control has become one of the main drivers of the burgeoning field of nanophotonics~\cite{Soukoulis2001book, Novotny2006book, Lourtioz2008book, Ghulinyan2015book, Barnes2019tutorial}. 
Following the seminal predictions by Bykov and by Yablonovitch, emission control was first studied on photonic crystals~\cite{Bykov1972, Yablonovitch1987PRL}. 
Emission control has also successfully been pursued with many different nanophotonic systems and many different quantum emitters, for instance,  atoms and dye molecules in Fabry–-P\'erot microcavities~\cite{Goy1983PRL, deMartini1987PRL}, quantum dots in pillar microcavities~\cite{Gerard1998PRL, Pelton2002PRL}, ions in  whispering gallery-mode microspheres~\cite{Treussart1994OptLett, Mabuchi1994OptLett, Sandoghdar1996PRA}, 
dye molecules in plasmonic nanocavities and on nanoantenae~\cite{Anger2006PRL, Muskens2007NanoLett, Rose2014NanoLett, Hoang2015NatCom, Chikkaraddy2016Nature}, or dye in metamaterials~\cite{Noginov2010OptLett, Lu2014NatNano}. 
 
In the weak-coupling approximation in cQED that is also known as the Wigner-–Weisskopf approximation~\cite{Weisskopf1930ZPhys}, spontaneous emission of an excited quantum emitter is precisely described by Fermi's golden rule~\cite{Fermi1932RMP} wherein the radiative decay rate is linearly proportional to the local density of optical states (LDOS). 
The LDOS counts the available number of electromagnetic modes each weighted by their strength at each point $\mathbf{r}_0$ and the projection of their electric field along the axes $x$,$y$,$z$~\cite{Snoeks1995PRL,
Sprik1996EPL, Barnes1998JMO}. 
The LDOS depends sensitively on the close environment of the emitter. 
Interestingly, the LDOS not only controls spontaneous emission and blackbody radiation, but also plays a role in van der Waals and Casimir dispersion forces and in F\"orster resonant energy transfer between different emitters~\cite{Wubs2015NJP}. 
Since the LDOS represents the density of vacuum fluctuations, it controls the amount of vacuum noise experienced by a qubit~\cite{Clerk2010RMP}. 

From theory, it is well-known that the LDOS is radically inhibited at frequencies within the 3D band gap in an infinite three-dimensional (3D) photonic crystal~\cite{John1990PRL, Sprik1996EPL, Busch1998PRE, Li2001PRA, Vats2002PRA, Wang2003PRL, Nikolaev2009JOSAB,Economou2010Book}. 
The LDOS vanishes at any position in the unit cell, and thus throughout the whole crystal, as well as for all dipole orientations. 
Concerning photonic crystal experiments, the first studies were reported on 3D crystals without 3D band gap~\cite{Yamasaki1998APL, Yoshino1998APL, Blanco1998APL, Megens1999PRA, Lodahl2004Nature, Nikolaev2007PRB, Nikolaev2008JPCC, Vallee2007PRB, Ventura2008AdvMat, Vion2009JAP, Subramaniam2009APL, Jorgensen2011PRL, Ning2012AdvMat}, or on band gap crystals with low-efficiency emitters~\cite{Koenderink2002PRL, Ogawa2004Science}. 
Leistikow \textit{et al.} studied efficient quantum dots in inverse woodpile photonic band gap crystals and observed exponential time-resolved decay, typical of weak coupling~\cite{Leistikow2011PRL}. 
A $10 \times$ inhibited spontaneous emission rate was observed inside the band gap. 
Since the emission was averaged over many emitters, it was inferred that a single quantum dot at the center of the crystal would be up to $160 \times$ inhibited. 
To date, however, these results have not been interpreted by theory or numerical calculations. 

It is obvious that experimental studies and devices employ finite photonic band gap crystals as energy can radiate from the boundaries of the finite crystal. 
Consequently, states from the infinite surrounding vacuum, tunnel into the crystal\footnote{A finite physical system, such as a photonic band gap crystal, that is surrounded by an infinite vacuum, or a bath, has in a strict mathematical sense a so-called finite support.}, leading to a non-zero LDOS and DOS inside the band gap~\cite{Yeganegi2014PRB, Hasan2018PRL}. 
Therefore, it is natural to wonder how the LDOS in the gap depends on the position and orientation of the emitter inside the crystal? 
For two-dimensional (2D) photonic crystals, Asatryan \textit{et al.} found in numerical calculations that the LDOS decreases exponentially from the surface into the crystal~\cite{Asatryan2001PRE}. 
Hermann and Hess found a strong position and orientation dependence of spontaneous emission within the unit cell of an inverse opal and saw that the inhibition in the band gap is of the order of two magnitudes, even for relatively small crystals~\cite{Hermann2002JOSAB}. 
Kole reported an exponentially growing inhibition at the center of a spherical inverse opal photonic band gap crystal~\cite{Kole2003thesis}. 
Leistikow \textit{et al.} proposed that the LDOS averaged over a unit cell decreases exponentially with position for frequencies inside the 3D band gap, with a characteristic length scale, the so-called LDOS decay length, but no prediction was offered for the dependence within the unit cell~\cite{Leistikow2011PRL}. 

Thus, it appears that calculations of the 3D LDOS in a 3D photonic band gap crystal are scarce in literature, due to their extensive computational cost and complexity. 
Therefore, we systematically investigate in this work the position and orientation-dependent inhibition of the LDOS in the band gap of 3D inverse woodpile crystals with finite support. 
Despite recent progress on analytical approaches in nanophotonics~\cite{deLasson2013JOSAB,Skipetrov2019arxiv}, there are to date no known analytic solutions for realistic 3D crystals, hence we have embarked on a numerical study to address the questions above. 
We study the role of the position and interpret the computational results by an analytical expression for the expected behavior of the LDOS. 
We also study the role of the dipole orientation, and compare it to theoretically known behavior~\cite{Vos2009PRA}.
Since we decided to investigate the experimental results of Leistikow \textit{et al.}~\cite{Leistikow2011PRL}, we have chosen to study the inverse woodpile crystal structure that was originally proposed by Ho \textit{et al.}~\cite{Ho1994SSC} 
In our study, we find remarkable physical features, namely, (1) that the LDOS decreases exponentially with position in the crystal, (2) the magnitude of the exponential length scale, the LDOS decay length $\ell_{\rho}$, is mostly determined by far-field radiation effects whereas the amplitude prefactor is mostly determined by near-field effects,
and (3) the magnitude of the LDOS decay length $\ell_{\rho}$ is remarkably close to the Bragg length - that typefies directional transport~\cite{Lin1992PRL, Vlasov1997PRB, NeveOz2004JAP} - which implies that the LDOS is strikingly directional. 

\section{Methods}
\label{sec:methods} 
\subsection{The structure of the finite crystal}
\label{sec:structure}

\begin{figure}[h!]
\centerline{ 
\includegraphics[width=8.6cm]{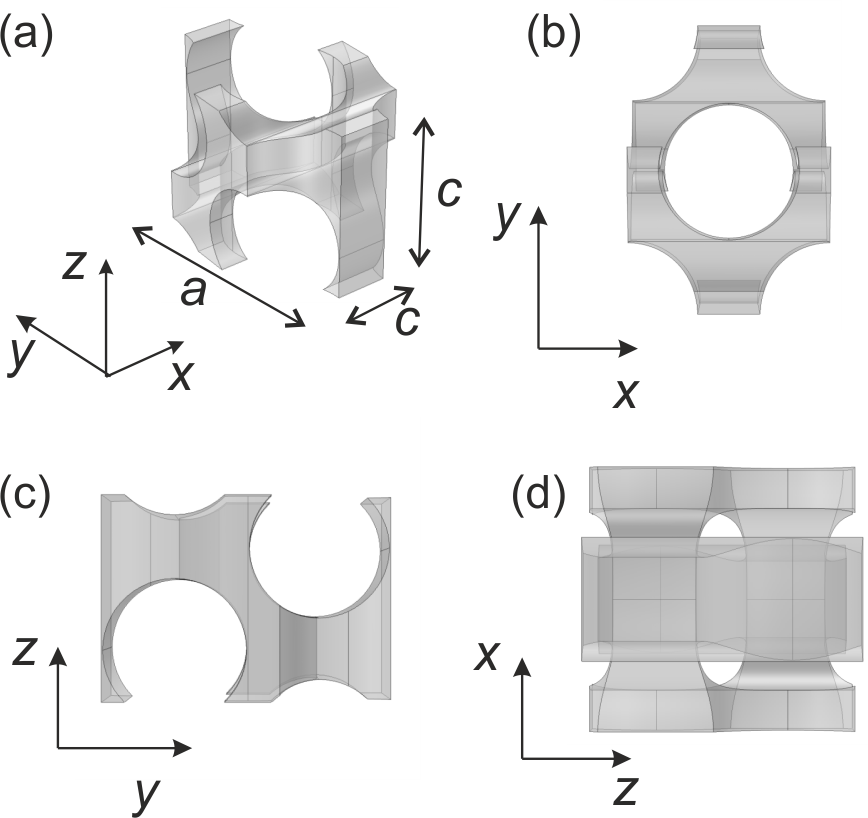}}

\caption{
Unit cell of the inverse woodpile structure.
(a) Bird's eye view of the tetragonal unit cell with two cylinders per lattice point with lattice parameters $c$ in the $x$-direction, $a$ in the $y$-direction, and $c$ in the $z$-direction. 
(b) View of the $xy$ face of the unit cell, (c) of the $yz$ face, and (d) of the $xz$ face. 
}
\label{fig:woodpile}
\end{figure} 
The inverse woodpile photonic crystal has a primitive unit cell that is illustrated in Figure~\ref{fig:woodpile}. 
The crystal structure consists of two orthogonal 2D arrays of identical cylindrical pores with radius $r_{p}=0.24\text{a}$ running parallel to the $x$ and $z$ axes~\cite{Ho1994SSC}. 
The lattice constants are $\text{a}$ (in the $y$-direction) and $c$ (in the $x$ and $z$ directions) in a ratio $\text{a}/ c = \sqrt{2}$ for the crystal structure to be cubic with a diamond-like symmetry. 
We discuss the LDOS as a function of the reduced frequency $\tilde{\omega}$ that is defined as $\tilde{\omega} \equiv \omega \text{a}/(2 \pi c_0)$ with $c_0$ the speed of light in vacuum. 
The backbone of the crystal has the dielectric constant $\varepsilon_b = 12.1$, typical of silicon in the near infrared and telecom spectral ranges. 
The cylindrical pores are considered to be either empty ($\varepsilon_p = 1$) or filled with a dielectric with $\varepsilon_p = 2.25$ that is typical for liquids such as toluene that are used to suspend quantum dot emitters in experiments, see Ref.~\cite{Leistikow2011PRL}. 
In the experimentally relevant spectral range, silicon and toluene are essentially lossless.
The finite crystals have an extent of $N$ unit cells along each of the $x$, $y$, and $z$ axes with a total volume of $V = N^3$ unit cells. 

\begin{figure}[t!]
\centerline{ \includegraphics[width=8cm]{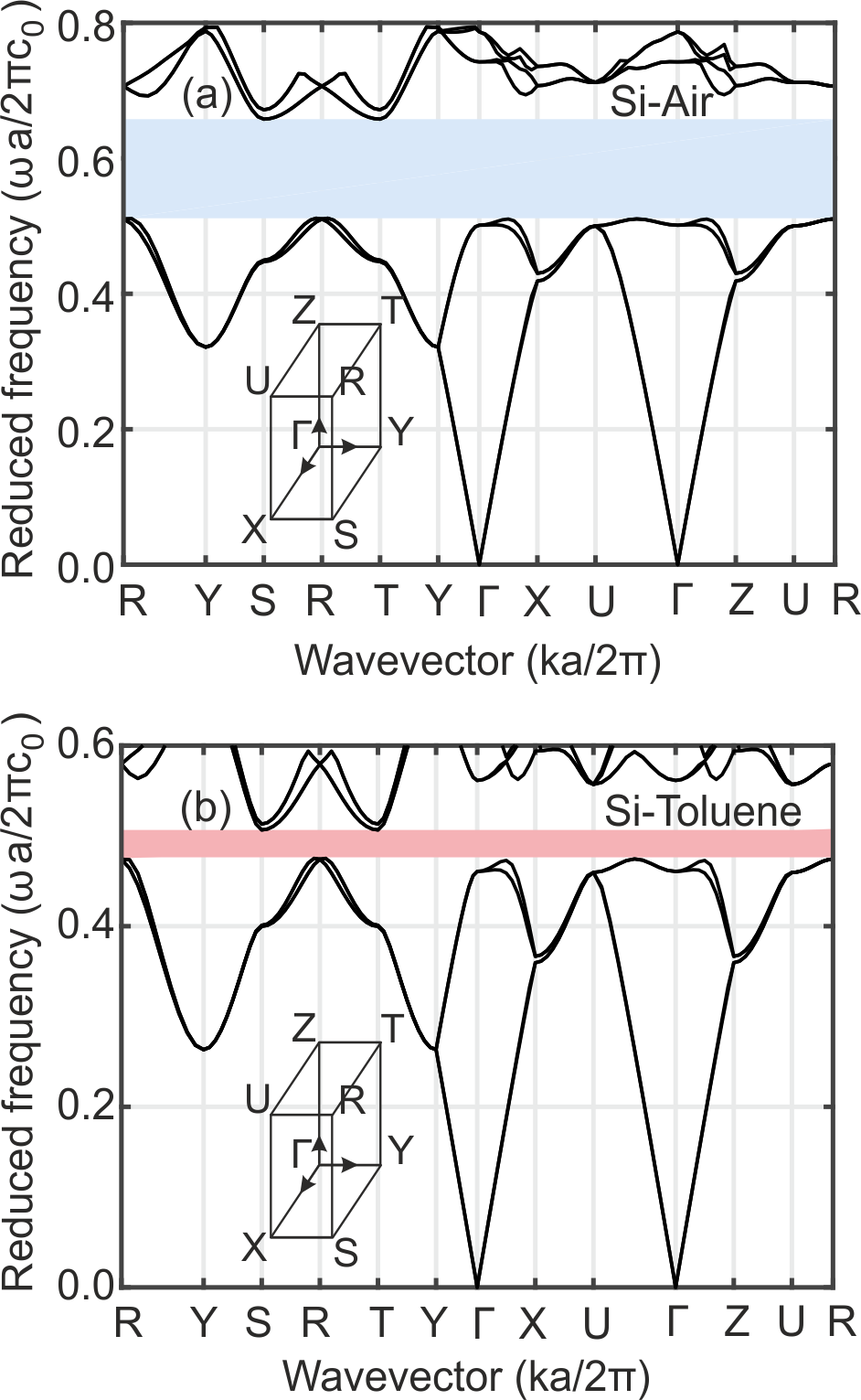} }
\caption{
Band structures for an infinite inverse woodpile crystal made of silicon ($\varepsilon_b=12.1$) with cylindrical pores filled with (a) air ($\varepsilon_p=1.0$) and (b) toluene ($\varepsilon_p=2.25$). 
The letters on the $x$-axis stand for the high symmetry points of the Brillouin zone shown in the inset. 
The blue and pink shaded bars indicate the 3D photonic band gap, from $0.511$ to $0.658$ and from $0.475$ to $0.507$, respectively.
}
\label{fig:woodpile_bands}
\end{figure} 

Figure~\ref{fig:woodpile_bands}(a) shows the band structure of the infinite crystal with empty pores calculated using the plane wave expansion method~\cite{MPBJohnson2001OptExpress}. 
The shaded area in Figure~\ref{fig:woodpile_bands}(a) indicates the 3D photonic band gap with a broad relative bandwidth $\Delta \tilde{\omega}/\tilde{\omega}_{\text{mid}} = 25.0\%$ centered at $\tilde{\omega}_{\text{mid}} = 0.585$, in good agreement with earlier work~\cite{Hillebrand2003JAP, Woldering2009JAP, Devashish2017PRB}. 
Figure~\ref{fig:woodpile_bands}(b) shows the band structure for the crystal filled with toluene. 
Due to the decreased dielectric contrast the 3D photonic band gap has a reduced relative width $\Delta\tilde{\omega}/\tilde{\omega}_{\text{mid}}=6.4\%$.
The band gap is centered at a lower frequency near $\tilde{\omega}_{\text{mid}} = 0.49$ due to the increased effective average dielectric constant~\cite{Datta1993PRB}. 
It is seen that in both air- and toluene-crystal cases the $\Gamma$Y stopgap is larger than the $\Gamma$X and the  $\Gamma$Z.
This is sensible since in the $y$-direction one encounters dielectric contrast of two set of pores whereas along the $x$- and $y$-direction one encounters dielectric contrast for only one set of pores.  

\subsection{Computation of the local density of states}
\label{sec:LDOS-calculation}
It is well-known that the LDOS $\rho^{(i)}(\omega,\mathbf{r}_0)$ at a point $\mathbf{r}_0$ projected along the $i$-axis ($i = \lbrace x,y,z\rbrace$) is proportional to the total power $P^{(i)}(\omega,\mathbf{r}_0)$ radiated by an electric point dipole current source  $\mathbf{J}(\omega,\mathbf{r}_0)=-i\omega\mathbf{p}(\omega)\delta(\mathbf{r}-\mathbf{r}_0)$ with dipole moment $\mathbf{p}(\omega)$ that points along the unit vector $\mathbf{\hat{e}}_i$ of the $i$-axis. 
\footnote{In \texttt{MEEP} the current source is defined as $\mathbf{J}(\omega,\mathbf{r}_0) =  \mathbf{p}(\omega)\delta(\mathbf{r}-\mathbf{r}_0)$  and hence, the power is: $P^{(i)}(\omega,\mathbf{r}_0)=-(1/2) \mathtt{Re}[\mathbf{E}\cdot \mathbf{p}^*(\omega)]$ }
It is therefore convenient to normalize the total power emitted inside a nanostructured medium to the power $P_0^{(i)}(\omega,\mathbf{r}_0)$ emitted by a same dipole in a homogeneous isotropic medium with the same dielectric constant $\varepsilon$ as where the dipole sits in the nanostructure. 
The normalized power is equal to the ratio of the LDOS in the nanostructured medium and the LDOS in a homogeneous medium with dielectric constant $\varepsilon$, and reads~\cite{Oskooi2013Chapter4}: 
\begin{equation}
\frac{\rho^{(i)}(\omega,\mathbf{r}_0)}{\rho_{0}^{(i)}(\omega,\mathbf{r}_0)} = \frac{P^{(i)}(\omega,\mathbf{r}_0)}{P_0^{(i)}(\omega,\mathbf{r}_0)}. 
\end{equation}
Using Poynting's theorem~\cite{Jackson1999Book}, the power $P^{(i)}(\omega,\mathbf{r}_0)$ radiated by the dipole at position $\mathbf{r}_0$ is equal to inner product of the dipole moment and the local electric field $\mathbf{E}(\omega,\mathbf{r}_0)$ at the position of the dipole
\begin{equation}
P^{(i)}(\omega,\mathbf{r}_0) = \frac{1}{2}\omega \mathtt{Im}\left[\mathbf{E}(\omega,\mathbf{r}_0)\cdot \mathbf{p}^*(\omega)\right]
\label{eq:power}
\end{equation}
where we use complex notation and consider steady-state (time-average).

To calculate the power radiated by the dipole inside the finite-size photonic crystals we used the open-source implementation \texttt{MEEP}~\cite{MEEP1} of the finite-difference time domain (FDTD) method~\cite{Yee1966IEEE}.  
The finite-size crystal is surrounded by a uniform dielectric buffer with the same dielectric constant as that of the low-$\varepsilon$ material in the pores. 
The computational volume is bounded on all sides by perfectly matched layers of thickness $\text{a}$ to emulate infinite space. 
A dipolar point source is placed at the position of interest $\mathbf{r}_0$ with a Gaussian spectrum with a central frequency equal to the mid-gap frequencies $\tilde{\omega}_\text{mid} = 0.58$ and $\tilde{\omega}_\text{mid} = 0.49$ for empty and toluene-filled crystals, respectively. 
The full width half maximum (FWHM) of the source spectrum was chosen to be equal to $\Delta \tilde{\omega} = 0.8$  to cover all the spectral features of interest. 

To assess possible numerical artifacts of our method, we have compared the computed LDOS at the center of a dielectric Mie-sphere with analytical results~\cite{Chew1987JChemPhys}, where the details are presented in the Appendix~\ref{app:validation}. 
For the best resolution (smallest grid size) and for a frequency range around the central frequency of the Gaussian pulse, we find convergence up to $3\%$ outside Mie resonances, and about $10\%$ near Mie-resonances as shown in Figure~\ref{fig:sphere-ldos}. 
The spatial grid size of $\Delta = \text{a}/30$ was used in the photonic crystal calculations, since this gave the best match with the analytic test results for a Mie sphere (see Appendix~\ref{app:validation}), while keeping the computation time within reasonable bounds.
The calculations were performed on a workstation with an Intel Core i7 processor with 8 CPU cores at 3.4 GHz clock speed and with 32GB RAM. 
To keep the simulations tractable, we studied 3D finite crystals with a volume $V = N^3 = 3^3$ unit cells.
The simulation times were equal to $600 (\text{a}/c_0)$, the real computation time was around $5000$ s in order to achieve sufficient convergence of our calculations. 

\section{Results and discussion}
\label{sec:results} 
\subsection{Local density of states versus emitter position}
\label{subsec:LDOS_vs_position}
\begin{figure}[t!]
\centerline{\includegraphics[width=8cm]{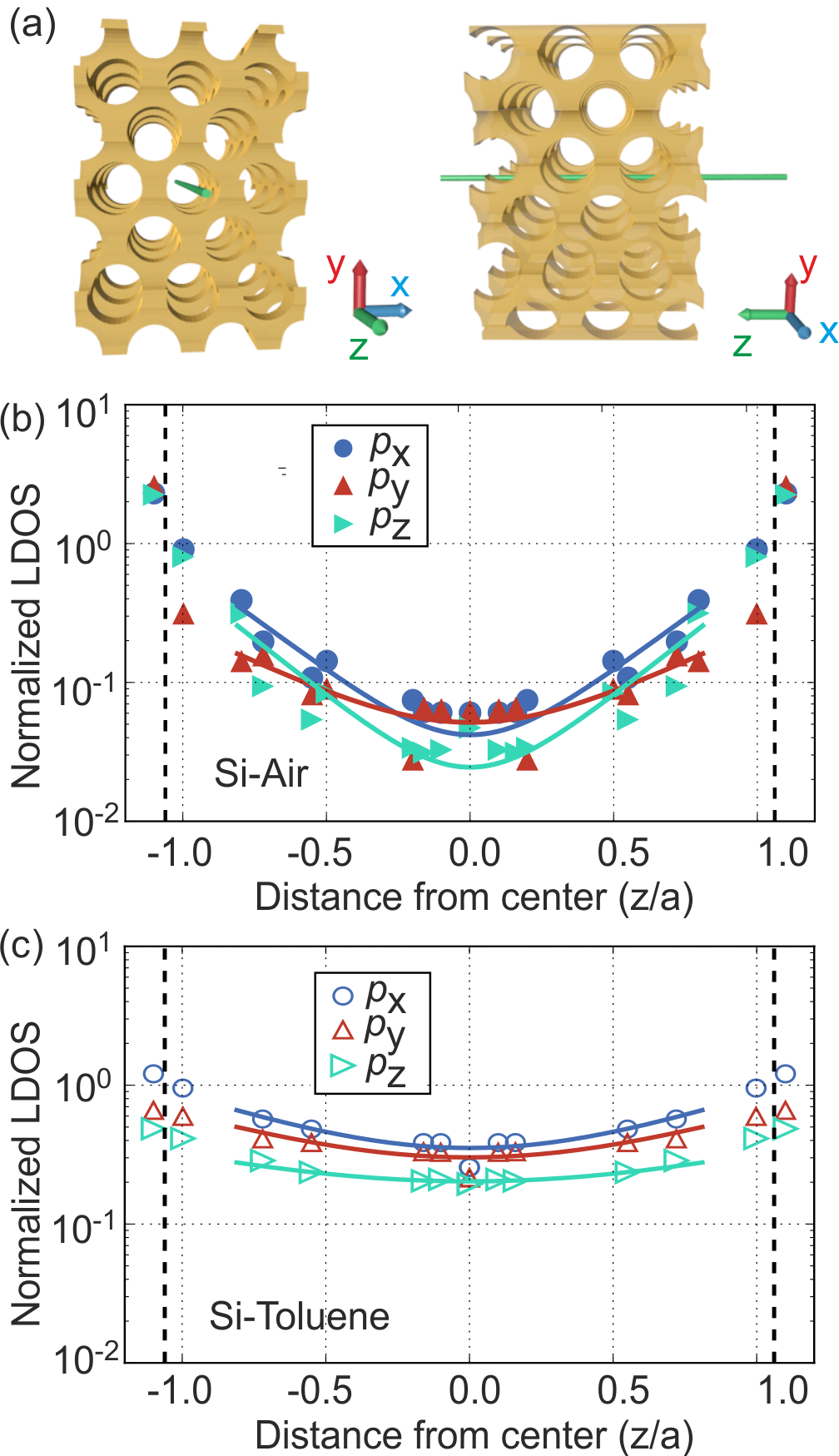}}
\caption{\label{fig:airSi_three_zaxis}
(a) Schematic of the $V = N^3 = 3^3$ crystal from two different perspectives, left: (001) view, right: (100) view. 
The green line ($x=0$, $y=0$, $z$) connects the positions where the LDOS is probed. 
(b, c) Normalized LDOS as a function of position along the $z$-axis at the mid-gap frequency for a (b) silicon-air crystal ($\tilde{\omega}_{\text{mid}} = 0.58$)  and (c) a toluene-air crystal ($\tilde{\omega}_{\text{mid}} = 0.49$)  with size $N^3=3^3$.
Blue circles are for x-dipoles $p_x$, red up-pointing triangles for y-dipoles $p_y$ and green right-pointing triangles for z-dipoles $p_z$. 
The drawn curves are exponential models of the data [Eq.~(\ref{eq:rho_vs_rFit})], with colors matching the relevant dipole orientation. 
The extent of the crystal is indicated by vertical dashed lines.
}
\label{fig:three_zaxis_midgap}
\end{figure}
We turn to the dependence of the LDOS $\rho^{(i)}(\omega,\mathbf{r}_0)$ on the position $\mathbf{r}_0$ inside the crystal at frequencies $\omega$ inside the 3D photonic band gap. 
We study the LDOS along trajectories in three different high-symmetry directions, where we make sure that all trajectories are centered. 
First, we consider the LDOS along the axis of the central pore pointing in the $z$-direction, as shown in Figure~\ref{fig:three_zaxis_midgap}. 
Figure~\ref{fig:three_zaxis_midgap}(a) illustrates the $(x=0, y=0, z)$ positions where the LDOS is probed. 
This set of probe positions are all in the same embedding medium (either air or toluene), which facilitates the interpretation. 
Figure~\ref{fig:three_zaxis_midgap}(b) presents the calculated LDOS for the silicon-air crystal at the mid gap frequency $\tilde{\omega}_{\text{mid}} = 0.58$ and Figure~\ref{fig:three_zaxis_midgap}(c) shows the LDOS for the toluene-filled crystal at the mid gap frequency ($\tilde{\omega}_{\text{mid}} = 0.49$). 
The silicon-air data strongly decrease from the crystal surface to the center of the crystal. 
For $x$- and $z$-oriented dipoles, the normalized LDOS tends from about $1$ to $5 \cdot 10^{-2}$, corresponding to a relative inhibition of $20     \times$ at the center. 
For the $y$-oriented dipole, the normalized LDOS tends from about $0.4$ to $5 \cdot 10^{-2}$, corresponding to a relative inhibition of $8 \times$ at the center. 
In the toluene-filled crystal, see Fig.~\ref{fig:three_zaxis_midgap}(c), similar trends appear, although with smaller inhibitions of about $2 \times$ to $3 \times$, since the refractive-index contrast and thus the photonic strength is less than in the air-filled crystal. 
Aside, we note that the LDOS near the crystal surface is slightly enhanced (for $x$- and $z$-oriented dipoles) or slightly decreased for $y$ dipoles, which we tentatively attribute to surface modes~\cite{Joannopoulos2008book} or to the fact that the vacuum modes are reflected by the crystal surface thus leading to interference just outside the surface, similar to the well-known Fresnel interference just outside a mirror~\cite{Jackson1999Book}. 

Let us briefly compare to the experiments by Leistikow \textit{et al.}~\cite{Leistikow2011PRL}, who studied the emission of quantum dot nanocrystals suspended in toluene that were embedded in silicon inverse woodpile structures. 
In the corresponding Fig.~\ref{fig:three_zaxis_midgap}(c), we observe a substantial inhibition of the LDOS, in agreement with the experimental observations. 
In the current situation, the inhibition at the center of the crystal is less (2 to 8 times) than in the experiments (more than 10 times), which is sensible since in the present case the crystal is smaller ($N^3 = 3^3$) than the ones in the experiments ($N^3 = 12^3$). 
There are aspects where no definite statements can be made, for instance, since the current results pertain to a single dipole that has a definite orientation, whereas in the experiment an ensemble of quantum dots was studied that sampled many positions throughout the whole crystal ($80 \%$ of the whole volume) and whose dipole orientations were random. 

Since the trend of the LDOS versus $z$-position in Figure~\ref{fig:three_zaxis_midgap}(b) is exponential within the domain that is computationally tractable here, we interpret the data with a model consisting of two exponentials:
\begin{equation}
\frac{\rho^{(i)}(z)}{\rho_0} = A_i( e^{z/\ell_{\rho}^{(i)}} + e^{-z/\ell_{\rho}^{(i)}}). 
\label{eq:rho_vs_rFit}
\end{equation}
The main characteristic is the LDOS decay length $\ell_{\rho}^{(i)}$ that pertains to dipole orientation $\mathbf{\hat{e}}_i$. 
In case of a strong inhibition of the LDOS, as is the case in a broad 3D photonic band gap, $\ell_{\rho}^{(i)}$ will be small, and $\ell_{\rho}^{(i)}$ increases for less inhibition. 
\footnote{Conversely, the inverse LDOS decay length may be considered to be a qualitative measure for the strength of a band gap.} 
As discussed below, the LDOS length $\ell_{\rho}^{(i)}$ is connected to far-field radiation effects of the dipole. 

In Eq.~(\ref{eq:rho_vs_rFit}) each exponential originates from one of the two opposite $(x,y)$-surfaces of the crystal (at $z/\text{a} = \pm 1.5/\sqrt2 = \pm 1.06$), hence the plus and minus signs with twice the same characteristic LDOS decay length $\ell_{\rho}^{(i)}$. 
And $A_i$ is a prefactor that equals half the LDOS at the center of the crystal (since $\rho^{(i)}(z=0)/\rho_0 = 2 A_i$). As discussed below, $A_i$ appears to be connected to near field effects of the dipole. 
In the modeling of the computed LDOS data with Eq.~(\ref{eq:rho_vs_rFit}), we exclude the two data points near the surface to avoid complications due to surface and edge states and Fresnel interference. 
The solid curves in Fig.~\ref{fig:three_zaxis_midgap}(b) and Fig.~\ref{fig:three_zaxis_midgap}(c) are the fitted curves according to Eq.~(\ref{eq:rho_vs_rFit}) for both silicon-air and silicon-toluene crystals and each of the three dipole orientations. 
The exponential model tracks the calculated LDOS data better in the toluene-filled crystal than the air-filled crystal, likely since in the former case the LDOS shows a weaker spatial dependence due to the reduced dielectric contrast, hence deviations are expected to be smaller. 
The resulting LDOS decay lengths and the prefactors are listed in Table~\ref{tab:Dlengths} for both air-filled and toluene-filled crystals.

\subsection{Model parameters and far-field and near-field}
\label{subsec:parameter_discussion}
\begin{table}[htbp]
\centering
\begin{tabular}{|c|c|c|c|c|}
\hline 
 & Air & Air & Toluene & Toluene\tabularnewline
\hline 

Orientation & $\ell_{\rho}^{(i)}/\text{a}$ & $A_{i}$ & $\ell_{\rho}^{(i)}/\text{a}$ & $A_{i}$\tabularnewline
\hline 
$\mathbf{\hat{e}}_x$ & 0.286 & 0.021 & 0.653 & 0.176 \\ \hline
$\mathbf{\hat{e}}_y$ & 0.449 & 0.026 & 0.743 & 0.151 \\ \hline
$\mathbf{\hat{e}}_z$ & 0.267 & 0.012 & 0.973 & 0.101 \\ \hline
\end{tabular}
\caption{\label{tab:Dlengths} Parameters of Eq.~\eqref{eq:rho_vs_rFit} to model the normalized LDOS versus position along the $z$-direction shown in Fig.~\ref{fig:three_zaxis_midgap}(b) and (c) for crystals with $N  = 3$. 
Here, $\ell_\rho^{(i)}$ is the LDOS decay length and $A_i$ is the amplitude prefactor.
Parameters are given for silicon-air and for silicon-toluene crystals, and the rows are for dipoles oriented in the $x$, $y$, and $z$ directions. }
\end{table}

Table~\ref{tab:Dlengths} shows that for the air-filled crystal the LDOS decay lengths are consistently smaller than for the toluene-filled crystals for all dipole orientations $\mathbf{\hat{e}}_x$, $\mathbf{\hat{e}}_y$, and $\mathbf{\hat{e}}_z$. 
The shorter LDOS decay lengths are a direct consequence of the higher dielectric contrast in the air-filled crystal, which results in a broader gap (see Fig.~\ref{fig:woodpile_bands}) and thus stronger inhibitions. 
In their study on silicon-toluene crystals, Leistikow et al.~\cite{Leistikow2011PRL} inferred the LDOS decay length to be equal to about $\ell_{\rho}^{(i)}/\text{a} = 1$. 
This is in fair agreement (between $3$ and $35 \%$ greater) with the results in Table~\ref{tab:Dlengths}, which is a gratifying consistency between the experimental and computed results. 

\begin{figure}[h!]
\centerline{ 
\includegraphics[width=8cm]{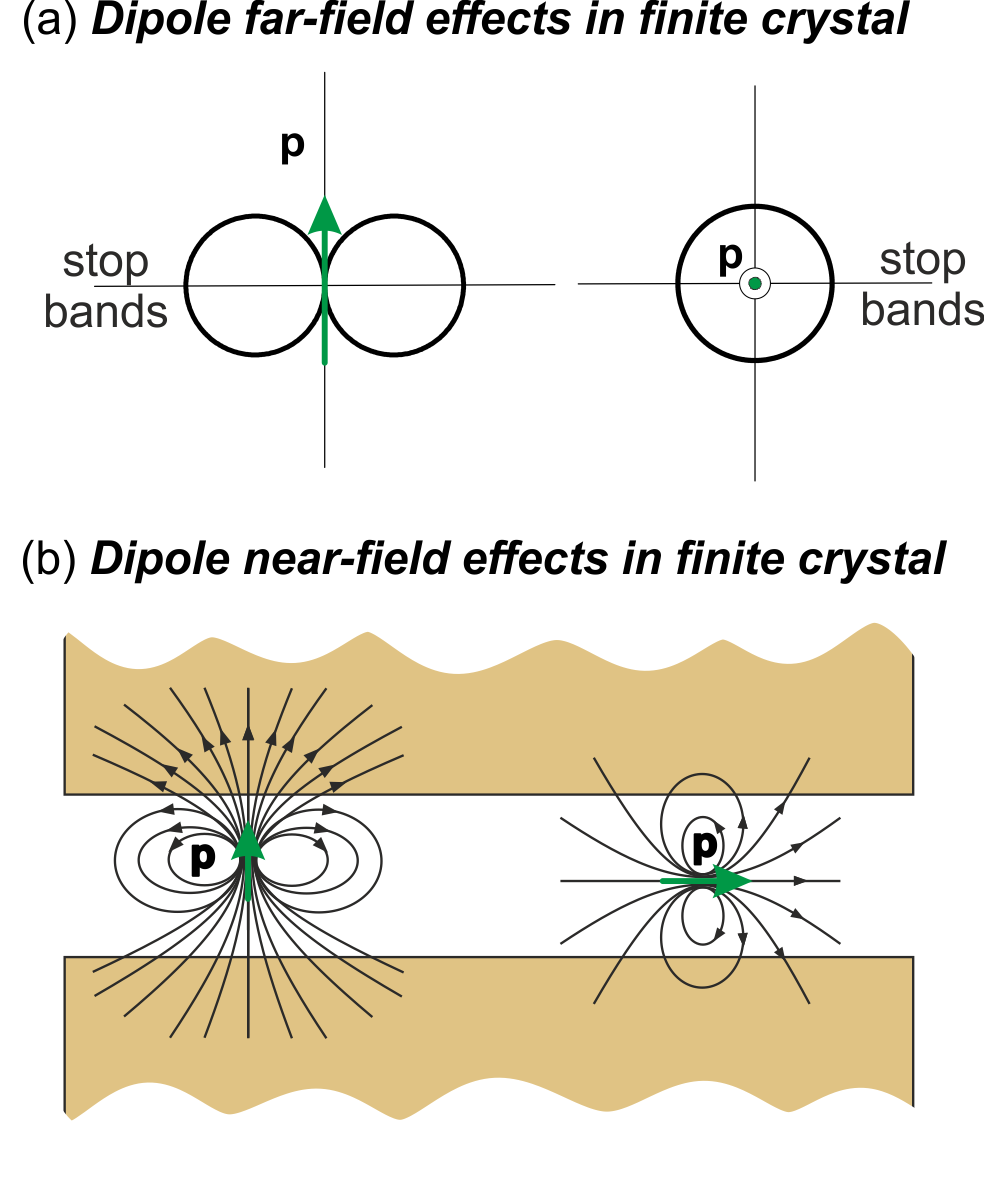}}
\caption{Schematic of a dipole (green) and its far-field radiation pattern  inside a finite photonic band crystal. 
(a) The far-field components are maximal in the equatorial plane, where stop bands affect the local density of states.
(b) For a dipole at the center of a pore, the near-field component is enhanced by the nearby high-index medium when the dipole is orientated toward this medium. 
The near-field component is hardly enhanced when the dipole is oriented along the pore axis ($x$ or $z$). }
\label{fig:cartoonNearFarFields} 
\end{figure}

When considering all parameters in Table~\ref{tab:Dlengths}, it is instructive to discuss the role of the dipole orientation $\mathbf{\hat{e}}_i$ on both the characteristic LDOS length $\ell^{(i)}_\rho$ and the amplitude prefactor $A_i$. 
Starting with the air-filled crystal, we observe that the $\mathbf{\hat{e}}_x$ and $\mathbf{\hat{e}}_z$ oriented dipoles have smaller LDOS decay lengths 
($\ell_{\rho}^{(x)} = 0.286$, $\ell_{\rho}^{(z)} = 0.267$) 
than the $\mathbf{\hat{e}}_y$ oriented dipole 
($\ell_{\rho}^{(y)} = 0.449$). 
This result can be rationalized by a simple model wherein we consider a dipole to have a far-field radiation pattern typical of a homogeneous medium, namely predominantly in its equatorial plane~\cite{Jackson1999Book} see Fig.~\ref{fig:cartoonNearFarFields}(a).
Hence, the $\mathbf{\hat{e}}_x$ dipole radiates predominantly in the $yz$-plane in the crystal. 
The light that would propagate in this plane notably encounters the $\Gamma$Y and the $\Gamma$Z high symmetry directions where the gap is wider (and intermediate directions) as seen in section~\ref{sec:structure}.  
Hence, we naturally expect a strong inhibition in the $yz$-plane, which agrees qualitatively with the small LDOS decay length for the $\mathbf{\hat{e}}_x$ orientation. 
A similar reasoning holds for the $\mathbf{\hat{e}}_z$ dipole, whose equatorial plane is the $xy$-plane in the crystal that again includes the $\Gamma$Y stop gap, and thus the LDOS decay length is also small.
Conversely, in case of the $\mathbf{\hat{e}}_y$ dipole, the equatorial plane is the $xz$-plane in the crystal. 
This plane contains the relatively narrower $\Gamma$X and $\Gamma$Z stop gaps (but not the broad $\Gamma$Y gap). 
Hence, less inhibition is expected than for the other orientations, which agrees well with the observed longer LDOS decay length. 
Thus, we conclude  that arguments based solely on the far-field radiation pattern of the dipole located within the photonic crystal serve to explain the relative strength of the characteristic LDOS length observed for different dipole orientations. 

We now turn to the role of the dipole orientation on the prefactor $A_i$. 
Here, we observe that the $\mathbf{\hat{e}}_z$ dipole exhibits the smallest prefactor ($A_z = 0.012$), whereas the  $\mathbf{\hat{e}}_x$  and the $\mathbf{\hat{e}}_y$ dipoles have almost twice greater  and closely similar prefactors ($A_x = 0.021$ and $A_y=0.026$). 
To understand this behavior, we recall that in the near-field regime a dipole has the strongest field component $\mathbf{E}(\omega,\mathbf{r}_0)_i$ in the same direction $i$ as its orientation $\mathbf{\hat{e}}_i$~\cite{Jackson1999Book}, as illustrated in Fig.~\ref{fig:cartoonNearFarFields}(b). 
Let us first consider the $\mathbf{\hat{e}}_y$ dipole orientation that has the maximum field in the $y$-direction $\mathbf{E}(\omega,\mathbf{r}_0)_y$.  
In the $y$-direction the $\mathbf{E}(\omega,\mathbf{r}_0)_y$ field crosses the air-silicon interface within a short distance equal to $\Delta y = 0.12 a = 0.12 \times 0.585 \lambda = \lambda / 14$. 
Therefore, this near-field experiences a polarization in the high-index material that enhances the near field. 
The enhanced near-field is apparently scattered to far-field radiation (by the interface), which leads to an enhanced LDOS and thus a larger prefactor $A_y$.  
Conversely, the $\mathbf{\hat{e}}_x$ and $\mathbf{\hat{e}}_z$ dipoles exhibit near fields in the $x$ and $z$-directions where the field is completely inside the uniform air-filled pore. 
Thus the concomitant near fields experience no polarization enhancement by the silicon, hence the smaller values of $A_x$ and $A_z$ prefactors.
Based of this reasoning we conclude that the $A_i$ prefactors are mainly associated with the near-field distributions of the $\mathbf{\hat{e}}_i$ oriented dipoles. 
For convenience, the results of our discussion are summarized in Table~\ref{tab:discussion}. 

\begin{table}[htbp]
\centering

\begin{tabular}{|c|c|c|}
\hline 
 & Far field & Near field \tabularnewline
\hline 
$p_x$  & More inhib. ($\Gamma Y$ stop gap) & Strong field  ($\perp$ diel.) \\ \hline
$p_y$  & \pbox{7cm}{Less inhib. \\ only $\Gamma X$,$\Gamma Z$ stop gaps} & Strong field  ($\perp$ diel.) \\ \hline
$p_z$  & More inhib. ($\Gamma Y$ stop gap) & Weaker field  ($\parallel$ diel.) \\ \hline
 & Concl.: affects $\ell_{\rho}^{(i)}$ & Concl.: affects $A_{i}$ \\ \hline\end{tabular}

\caption{\label{tab:discussion} 
Table summarizing the discussion of the near and far field effects on the LDOS decay lengths $\ell_{\rho}^{(i)}$ and the prefactor $A_{i}$. 
The LDOS decay lengths are mainly affected by the far field and the stop gaps in the directions of the radiation.
The prefactors $A_{(i)}$ are mainly affected by near field effects and by polarization effects due to presence of nearby high-index material. }
\end{table}

In the case of toluene-filled crystal, the behavior of the LDOS amplitude $A_i$ is similar as in the silicon-air crystal.
As seen in Table~\ref{tab:Dlengths}, the dipole with polarization $\mathbf{\hat{e}}_z$ exhibits the smallest amplitude $(A_z = 0.101)$, followed by the dipoles polarized along $\mathbf{\hat{e}}_y$ and $\mathbf{\hat{e}}_x$ that have similar amplitudes ($0.151$ versus $0.176$). 
We thus conclude that the near field has the same impact in this case. 
The characteristic LDOS length $\ell_\rho^{(i)}$, however, does not exhibit the same pattern as in the silicon-air case.
As shown in  Table~\ref{tab:Dlengths} the strongest inhibition in this case, that is, the smaller LDOS length is found for the  $\mathbf{\hat{e}}_x$ oriented dipole, followed by the $\mathbf{\hat{e}}_y$ dipole, and the $\mathbf{\hat{e}}_z$ dipole. 
The mismatch between the air and toluene cases is possibly caused by the fact that in the silicon-toluene crystal, the reduced refractive index contrast probably leads to an increase of the directional Bragg length to be larger than the crystal size of $3\times 3\times 3$ unit cells considered here.
In the toluene case, it is probably not meaningful to interpret the LDOS inside the crystal with band structure features, since the infinite crystal is not reached in the computations. 
In the silicon-air crystal, the Bragg length is sufficiently small compared to the crystal size that the infinite crystal limit is effectively reached, hence a reasoning invoking the interference associated with the stop gaps in the band structure is meaningful.

\subsection{Comparison between LDOS decay length and Bragg length}
\label{subsec:LDOS_versus_Bragg}

\begin{figure}[h!]
\centerline{ 
\includegraphics[width=8cm]{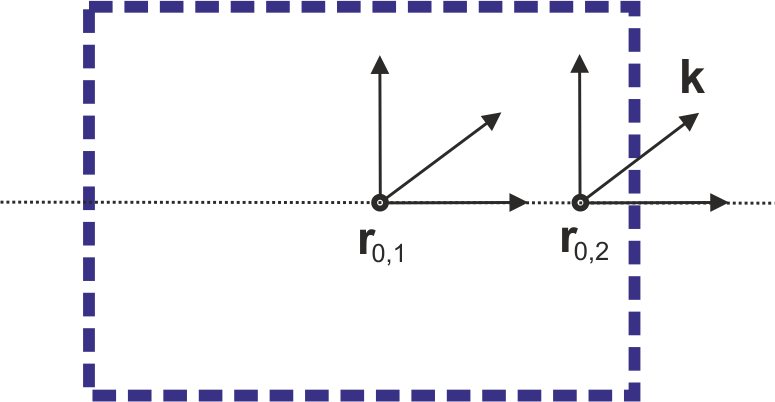}}
\caption{Schematic
representation of the position vector $\mathbf{r}$=$(r_x,r_y,r_z)$ and the complex wavevector  $\mathbf{k} = (k_x+ik_x',k_y+ik_y',k_z+ik_z')$ in the crystal under study comprising of $N = 3$ unit cells. 
Inside the band gap where we calculate the LDOS the imaginary part of $\mathbf{k}$ is nonzero. 
The position vector lies along the axis defined in each of the position dependence case studies in Fig.~\ref{fig:three_zaxis_midgap},  Fig.~\ref{fig:airSi_three_xaxis} and  Fig.~\ref{fig:airSi_three_xzdiag}.  }
\label{fig:cartoonLDOSvsdirection} 
\end{figure}

To put the LDOS decay length in perspective, we compare it to the well-known Bragg length $L_B$~\cite{Lin1992PRL, Vlasov1997PRB, NeveOz2004JAP} that describes the exponential decay of a directional incident light beam with a frequency inside a photonic gap. 
This directional decay is described by a nonzero imaginary part of the wavevector $\mathbf{k}''$ due to Bragg diffraction. 
The imaginary part of the wavevector is inversely equal to the Bragg length $L_B$. 

For silicon-air inverse woodpile crystals with the same pore radii as here, the Bragg length $L_B$ was computed by Devashish \textit{et }al.~\cite{Devashish2017PRB} by the finite-element method. 
For $x$-polarized incident plane waves, they found $L_B^{(x)} = 0.262\text{a}$, and $L_B ^{(y)}= 0.428\text{a}$ for $y$-polarized illumination. 
These values are similar to the LDOS decay lengths $\ell_{\rho}^{(x)} = 0.286\text{a}$ and $\ell_{\rho}^{(y)}= 0.449\text{a}$ in Table~\ref{tab:Dlengths} for $x$- and $y$-oriented dipoles, respectively. 
Considering the difference between the underlying physics, namely the LDOS versus directional propagation, in other words, the real part of the Green function~\cite{Economou2006Book} versus the imaginary part of the Green function, it is remarkable for the two different length scales to match so closely. 

To further support our interpretation, we consider in the schematic in Fig.~\ref{fig:cartoonLDOSvsdirection} a dipole at two different positions inside a finite crystal, where we assume the positions to be on the z-axis as in Fig.~\ref{fig:airSi_three_zaxis}. 
The dipole emits in many different directions in wave vector space (in wave vector space, a crystal as in Fig.~\ref{fig:airSi_three_zaxis} is more extended in the horizontal wave vector direction). 
Let us first consider an $\mathbf{\hat{e}}_x$-oriented dipole that radiates mostly in the $(y,z)$-plane. 
Since the dipole has a frequency within the band gap, the radiation in any direction will be exponentially damped, since the wave vector is in every direction complex. 
Thus, the radiation in the $z$-direction is less damped close to the crystal surface (at position $\mathbf{r}_{0,2}$) than deeper inside at position $\mathbf{r}_{0,1}$. 
Radiation in the perpendicular $y$-direction is equally damped at the different positions, since in this direction the dipole is everywhere at the same distance from the crystal-vacuum interface (similar considerations pertain to the $x$-direction). 
Radiation in an oblique direction with wave vector $\mathbf{k}$ will also be increasingly damped when the dipole is located at increasing depth in the crystal. 
The behavior seen in Fig.~\ref{fig:airSi_three_zaxis} suggests that apparently the behavior of the LDOS with $z$-position is mostly determined by the $z$-directed radiation (that is in its purest sense described by the Bragg length), and hardly by the $y$-directed or other oblique radiation with wave vector  $\mathbf{k}$. 
Thus, whereas the LDOS usually integrates over a broad spectrum of field modes with wave vectors corresponding to \textit{all} directions, apparently the radiation tending in the closest vacuum-crystal interface dominates the spectrum. 
On the other hand, the fact that the LDOS integrates over a broad spectrum, instead of a single wave vector as in the Bragg length, explains perhaps why the LDOS decay length is somewhat larger than the corresponding Bragg length.

\subsection{LDOS along different trajectories}
\label{subsec:LDOS_along_x_and_diagonal}

\begin{figure}[th!]
\centerline{\includegraphics[width=8cm]{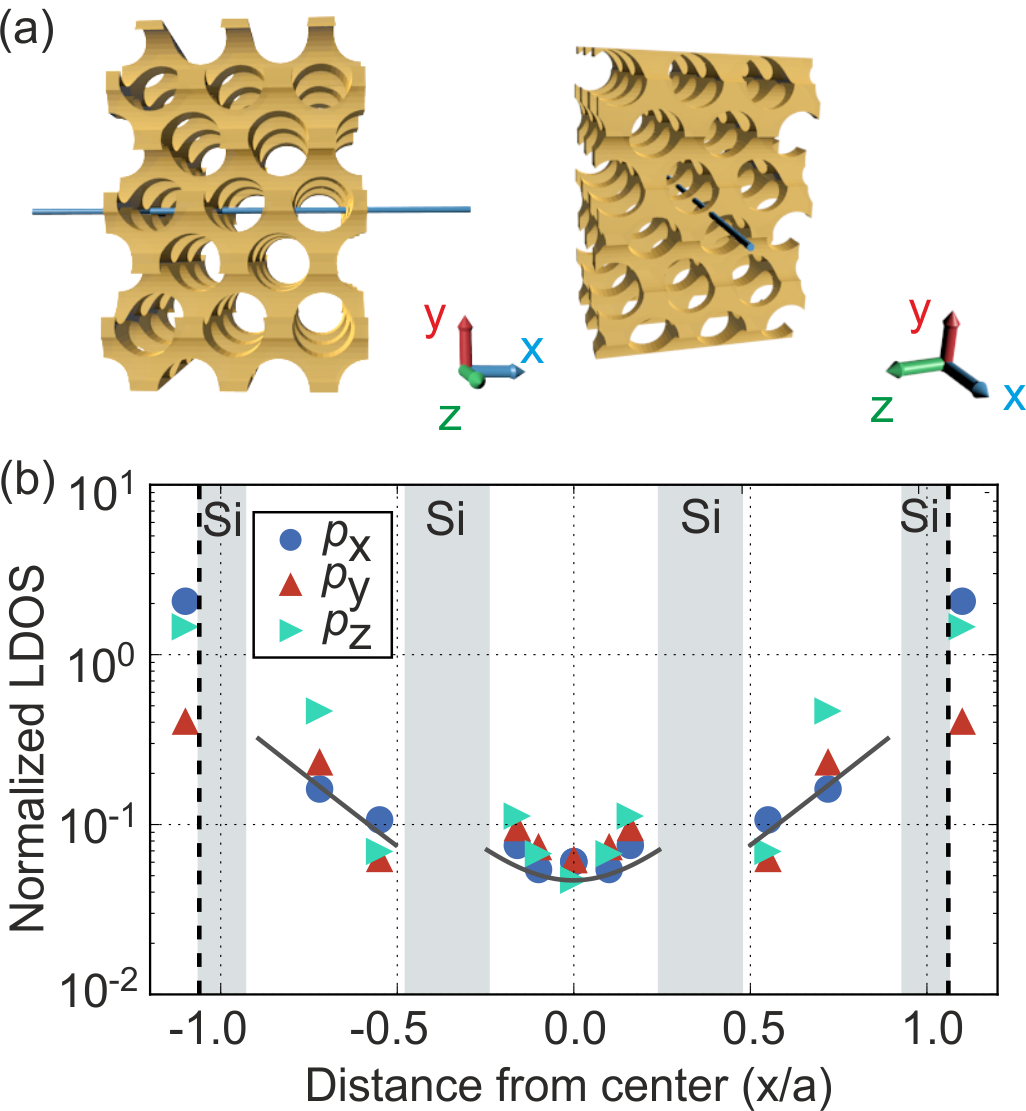}}
\caption{\label{fig:airSi_three_xaxis}
(a) Schematic of the $V = N^3 = 3^3$ crystal from two different perspectives, left: (001) view, right: (100) view. 
The blue line ($x$, $y=0$, $z=0$) connecting the positions where the LDOS is probed seen from two different perspectives.
(b) Normalized LDOS as a function of position along the $x$-axis at the mid-gap frequency ($\tilde{\omega}_{\text{mid}} = 0.58$) for a silicon-air crystal with size $N^3 =3^3$.
Blue circles are for x-dipoles $p_x$, red triangles are for y-dipoles $p_y$, blue-green right-pointing triangles are for z-dipoles $p_z$. 
The lines passing through the data points are guides to the eye. 
The shaded areas are the silicon backbone. 
}
\label{fig:three_xaxis_midgap}
\end{figure}

\begin{figure}[th!]
\centerline{\includegraphics[width=243pt]{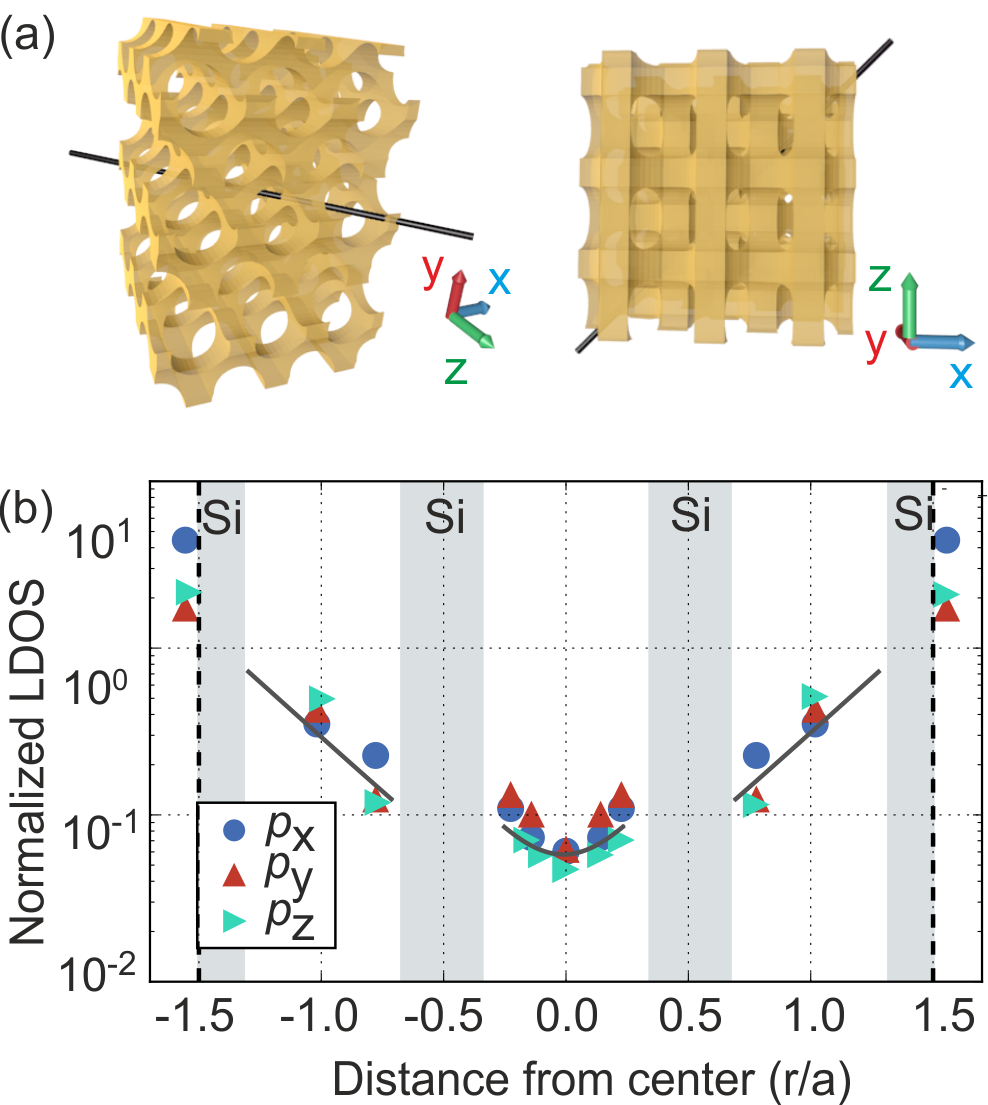}}
\caption{\label{fig:airSi_three_xzdiag} (a) Schematic of the $V = N^3 = 3^3$ crystal from two different perspectives, left: (001) view,  right: (0$\overline{1}$0) view. 
The black line ($x$, $y=0$, $z=x$) connecting the positions where the LDOS is probed. 
(b) Normalized LDOS as a function of position along the $xz$-diagonal ($y=0$) at the mid-gap frequency ($\tilde{\omega}_{\text{mid}} = 0.58$) for a silicon-air crystal with size $N^3 = 3^3$. 
Blue circles are for x-dipoles $p_x$, red triangles are for y-dipoles $p_y$, blue-green right-pointing triangles are for z-dipoles $p_z$. 
The lines passing through the data points are guides to the eye and the shaded areas are the silicon backbone. 
}
\end{figure}

We show in Figures~\ref{fig:three_xaxis_midgap} and \ref{fig:airSi_three_xzdiag} the normalized LDOS on the trajectories along $x$-axis and the diagonal path on the $xz$-plane, respectively. 
In both of these cases the calculations refer only to the silicon-air crystal.
Once again, the position of the emitters are shown in the schematics of the upper panels of  Fig.~\ref{fig:three_xaxis_midgap} and Fig.~\ref{fig:airSi_three_xzdiag} respectively while the LDOS values are plotted in the bottom panel.
Along these paths, the emitter goes through both the silicon and air regions inside the crystal. 
Since in the experiments by Leistikow \textit{et al.} the LDOS was only probed for emitters placed in the low-index region, we have not considered the LDOS inside the high-index silicon. 
While moving across the air-Si interface, the normalized LDOS does not reveal a smooth and continuous behavior, which makes it impossible to use a simple exponential model such as Eq.~\eqref{eq:rho_vs_rFit}. 
Indeed, similar strongly varying behavior across the low and high-index regions within a unit cell has already been noted in Ref.~\cite{Yeganegi2014PRB} for the much simpler case of a finite-size Bragg stack (aka, a "one-dimensional photonic crystal"). 

To highlight this behavior in our 3D crystal, we only draw guides to the eye that mark the trend of the LDOS in each direction. 
They are marked as black solid curves in  Fig.~\ref{fig:three_xaxis_midgap}(b) and Fig.~\ref{fig:airSi_three_xzdiag}(b). 
In both cases, it appears, that LDOS reveal abrupt variations while tending across the Si regions, which are highlighted in gray. 
Interestingly, for the $x$-polarization when moving towards the vacuum-crystal interface from the center, LDOS shifts down across the silicon region in the LDOS calculated on $x$-axis (Fig.~\ref{fig:three_xaxis_midgap}(b)) whereas it shifts up in the LDOS values calculated on the diagonal of the $xz$-plane (in Fig.~\ref{fig:airSi_three_xzdiag}(b)).

\subsection{Practical consequences}
\label{subsec:consequences}
Let us briefly discuss a number of practical implications of our work, namely how to apply 3D photonic band gaps to emission control, quantum information processing, and  photovoltaics. 

In the field of spontaneous control, since the radiative rate is proportional to the LDOS, controlling the LDOS is a key step~\cite{Barnes2019tutorial}. 
Hence it is clear that a 3D photonic band gap offers an extreme spontaneous emission control.
In the field of photovoltaics, it has been realized that an efficient absorber is equivalent to an efficient emitter~\cite{Polman2012NatMat}. 
Hence a 3D photonic band gap could offer a control means to photovoltaics.
In the field of quantum information science, it is relevant to shield qubits from ubiquitous vacuum fluctuations that lead to decoherence of the quantum states~\cite{Clerk2010RMP, Vos2015book}. 
One solution to this challenge is to place the qubits (assuming they are dipolar) in a 3D photonic band gap that covers the relevant frequency range of the qubits. 
Our work provides a design rule, namely where to place a dipolar emitter inside a photonic band gap crystal for a certain emission control, and equivalently, where to place a dipole for a certain absorption control, and again equivalently where to place a qubit for a certain decoherence control.

For instance, if one requires the density of vacuum fluctuations - hence the LDOS - to be shielded by a factor $10 \times$, Fig.~\ref{fig:three_zaxis_midgap} shows that this is feasible for dipoles placed anywhere between $-0.5\text{a} \leq z \leq +0.5 \text{a}$ about the center. 
For dipoles operating at optical frequencies corresponding to 1500 nm in the telecom range, this position freedom corresponds to a relatively large range of about 700 nm. 
A slight limitation to our study is that we only consider positions in the low-index medium of the photonic crystal nanostructure, although these positions occupy no less than $80$ vol$\%$ of the whole crystal volume~\cite{Devashish2017thesis}. 
The results in Fig.~\ref{fig:three_zaxis_midgap} also show that a tenfold shielding of the vacuum fluctuations is robust with respect to the orientation of the transition dipole moment of the dipole. 

It is exhilarating that a silicon-air crystal has a significant inhibition of the LDOS, in view of the relatively small crystal size of only $V = 3\times 3 \times 3$ unit cells. 
For dipolar emitters (qubits) operating at optical frequencies corresponding to 1500 nm, this would corresponds to a small 3D silicon nanophotonic device with a volume as small as $V = 4.2~\mu$m$^{3}$.  
Such a robustness with respect to the crystal size is due to the small LDOS decay length that is much less than one lattice spacing. 
In parallel to this paper, an experimental study of the directional stop bands of (necessarily finite) 3D photonic band gap crystals~\cite{Tajiri2020arxiv} is also reaching the conclusion that relatively small micron-sized crystals are powerful tools to control directional transport.

\section{Conclusions}
\label{sec:conclusions}
In this paper, we have presented a computational study of the inhibition of the LDOS in the 3D photonic band gap of a finite-size 3D photonic crystal. 
In particular, we focused on crystals with the silicon inverse woodpile structure that were recently studied experimentally. 
To this end, we considered the LDOS dependence on emitter's position and orientation inside the crystal. 
Our calculations showed that except for special cases, it is generally not possible to model the LDOS decrease away from the vacuum-crystal interface with a simple exponential model. However, where the exponential model did work, the LDOS decay length turned out to be surprisingly similar to the Bragg length. As for the impact of crystal size on the LDOS suppression, we found that a crystal only as large as comprising 3x3x3 unit cells and with good dielectric contrast (silicon-air) already provided more than ten times inhibition of the LDOS around its center. Therefore, for experiments designed to shield quantum systems from vacuum fluctuations, very small volume devices may well be sufficient to fulfill the design requirements on LDOS suppression.   

\section{Acknowledgments}
\label{sec:acknowledgments}
It is a great pleasure to thank Devashish for help, and Michael Sigalas (Patras), Ad Lagendijk, and Pepijn Pinkse for useful discussions. 
This work was supported by the European Research Council under ERC Advanced Grant no. 320081 (project PHOTOMETA), the European Union’s Horizon 2020 Future Emerging Technologies call (FETOPEN-RIA) under grant agreement no. 736876 (project VISORSURF), the General Secretariat for Research and Technology and the H.F.R.I. PhD Fellowship grant (GA. no. 4894) in the context of the action "1st Proclamation of Scholarships from ELIDEK for PhD candidates. 
We also acknowledge support from (formerly FOM) NWO-Shell program CSER, NWO-Physics program "Stirring of Light!", and the MESA+ Institute for Nanotechnology Applied Nanophotonics (ANP) division. Work at Ames Laboratory was supported by the US Department of Energy, Office of Basic Energy Science, Division of Materials Science and Engineering (Ames Laboratory is operated for the US Department of Energy by Iowa State University under contract No. DE-AC02-07CH11358.

\appendix
\section{Numerical calculation of the LDOS}
\label{app:validation}
\begin{figure}[th!]
\centerline{ 
\includegraphics[width=8cm]{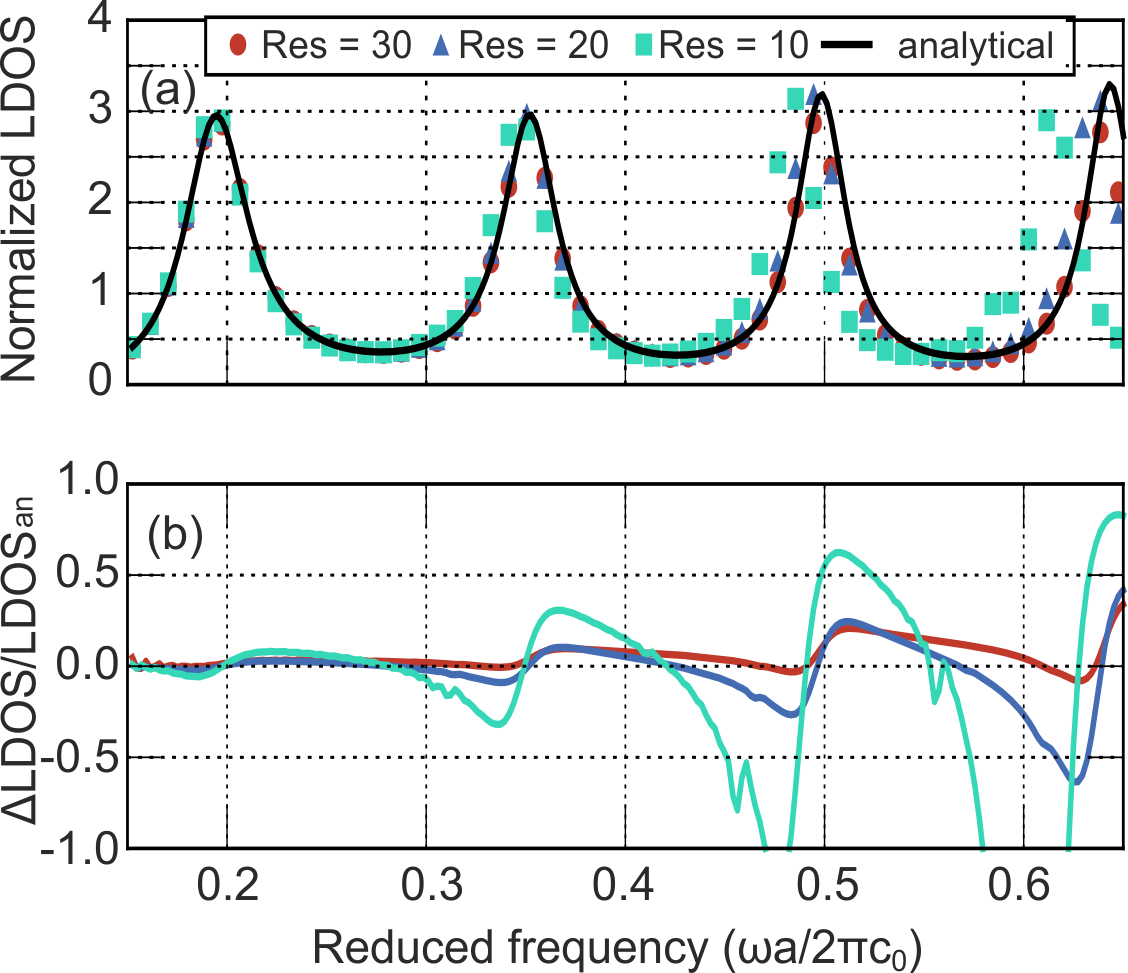}}
\caption{
\label{fig:sphere-ldos}
(a) LDOS at the center of a sphere for three different resolutions (10 (blue-green squares), 20 (blue triangles), and 30 points per radius (red circles)) of the FDTD grid versus the analytical solution (black curve). 
(b) Difference between the FDTD calculated results and the analytical solution for three different resolutions: 10 grid points per radius (blue-green squares), 20 grid points per radius (blue triangles), and 30 grid points per radius (red circles)). }
\end{figure} 
We numerically calculate the LDOS by relating the electric  field at the location of the electric point dipole to the power radiated by the dipole, see Eq.~(\ref{eq:power}). The electric field is obtained by placing a point dipole source at point $\mathbf{r}_0$ with a dipole moment parallel to $x$, $y$, or $z$ axes. 
The transient amplitude of the dipole moment is described by a short Gaussian pulse to generate sufficient band width to cover the entire frequency spectrum of interest.
After the initial excitation has vanished, we obtain the electric field component parallel to the dipole moment at $\mathbf{r}_0$ versus time $t$ at every time step and take the Fourier transform to obtain the frequency-resolved field $\mathbf{E}(\omega,\mathbf{r}_0)$. 
This approach has also been used in earlier studies too; see for instance Refs.~\cite{Xu1999JOSAB, Koenderink2006JOSAB, Hermann2002JOSAB, Ishizaki2009JOSAB}.

To validate the calculation of the LDOS with \texttt{MEEP} FDTD-code, we compare the results of FDTD method with analytical results, namely the modification of LDOS at the center of a dielectric sphere~\cite{Chew1987JChemPhys,Saunders1992JApplPhys}. 
We consider a sphere of radius $\textnormal{a} = 1$ (reduced units) made of a dielectric material with real dielectric constant $\varepsilon_b = 12.1$. 
Fig.~\ref{fig:sphere-ldos}(a) shows the LDOS predicted by exact calculations (solid line) which exhibits resonances at reduced frequencies $0.2$, $0.35$, $0.50$ associated with the Mie-resonances of the sphere. 
The numerical results were obtained with a Gaussian pulse centered at $\tilde{\omega}=0.4$ and width $\Delta\tilde{\omega}=0.9$. 
The numerically computed LDOS using the FDTD method is shown in the same figure as discrete points for various grid resolutions defined as the number of sampling points over a radial distance. 
Good agreement is found between the analytical and numerically computed LDOS specially at higher resolution (30 grid points).

In Fig.~\ref{fig:sphere-ldos}(b), we quantify the convergence between analytical and numerical results by showing the relative difference (in percentage) between the numerical and analytical results. 
We observe that the convergence is better at frequencies outside the resonances - up to $3\%$ near the central frequency $0.4$ of the spectrum - than for the frequencies around the resonances, typically up to $10\%$ near the central frequency $0.4$. 
The most extreme differences appear at the upper edge of the spectrum where the precision is limited by the low intensity of the excitation pulse in the computation. 
This is expected due to the fact that the lifetime of the modes on-resonance is much greater and hence, more computational time is required for these frequencies.
The numerical calculations are in an excellent agreement with the analytical results, to within $10\%$ near resonances and $3\%$ for off-resonance frequencies. 
Therefore, we conclude that the simulated results provide a faithful representation of the physics under study.

\end{document}